\documentclass[journal=apchd5,manuscript=letter]{achemso}

\usepackage[T1]{fontenc} 

\author{Dror Liran}
\altaffiliation{These two authors contributed equally}
\author{Itamar Rosenberg}
\altaffiliation{These two authors contributed equally}
\affiliation{Racah Institute of Physics, The Hebrew University of Jerusalem, Jerusalem 91904, Israel}
\author{Kenneth West}
\author{Loren Pfeiffer}
\affiliation{Department of Electrical Engineering, Princeton University, Princeton, New Jersey 08544,USA}
\author{Ronen Rapaport}
\affiliation{Racah Institute of Physics, The Hebrew University of Jerusalem, Jerusalem 91904, Israel}
\affiliation{The Applied Physics Department, 
The Hebrew University of Jerusalem, Jerusalem 91904, Israel}
\email{ronenr@phys.huji.ac.il}

\title[An \textsf{achemso} demo]
  {Fully guided electrically-controlled exciton polaritons}

\begin{document}

\begin{abstract}
We demonstrate two types of waveguide structures which optically confine exciton-polaritons in two dimensions and act as polaritonic channels. We show a strong optical confinement in an etched rectangular waveguide, that significantly increases the propagation distance of the polaritons and allow to direct them in curved trajectories. Also, we show low-loss optical guiding over a record-high of hundreds of microns which is combined seamlessly with electrical control of the polaritons, in a strip waveguide formed by electrically conductive and optically transparent strips deposited on top of a planar waveguide. Both structures are scalable and easy to fabricate and offer new possibilities for designing complex polaritonic devices. 
\end{abstract}

\section{Introduction}
Exciton-polaritons are quantum superpositions of light and matter resulting from the strong coupling between confined optical modes  and confined electronic excitations (excitons).  They have very small effective masses $\sim10^{-5}m_0$ and very large propagation velocities, on the one hand, while maintaining the ability to interact with each other, as well with external electric or magnetic fields on the other hand. This conjunction of properties induces effectively strong nonlinearities to the medium leading to "interacting photons". As such, polariton quasi-particles hold promise for realizations of light-based quantum devices with new functionalities. 

Typically, semiconductor microcavities (MC) are used as means for optical confinement of the photonic part of the polaritons. In these structures, several quantum wells (QWs) are positioned at the anti-nodes of the optical mode, which is confined between two distributed Bragg reflectors (DBRs). Recently, several proofs-of-concept of non-linear photonic devices based on MC-polaritons have been demostrated\cite{amo_excitonpolariton_2010,ferrier_interactions_2011,cristofolini_optical_2013, gao_polariton_2012,ballarini_all-optical_2013,sturm_all-optical_2014,nguyen_realization_2013,Delteil2018QuantumExciton-polaritons,Sanvitto2016TheDevices,Liew2008OpticalMicrocavities,menon_nonlinear_2010}.  MC-based polariton devices have however some drawbacks when considering the feasibility of large-scale polaritonic-based circuits. Firstly, DBRs are monolithically grown thick multi-layers, making the fabrication process of even the simplest lateral structures complex and prone to surface defects and imperfections. Secondly, even very thick high-quality DBRs have a finite reflectivity, causing a constant photon leak as polaritons propagate inside the sample. This limits polariton propagation lengths even with high quality mirrors \cite{Nelsen2013DissipationlessLifetime} to distances typically below 200$\mu$m \cite{Wertz2010SpontaneousCondensates,Liew2010Exciton-polaritonCircuits}. Together with the inherent tendency of MC-polaritons to relax to their lowest energy having a zero in-plane momentum (and thus low in-plane propagation velocities), these short propagation distances limits the ability to create extended optical circuits based on MC-polaritons.

Several years ago, it was demonstrated that strong coupling between excitons in QWs and propagating optical modes of a planar slab waveguide (WG) leads to the formation of propagating ('flying') WG-polaritons\cite{walker_exciton_2013}. Since the optical modes in a WG are confined by total internal reflection, the optical losses are significantly reduced, and the WG-polaritons display very high propagation velocities and long propagation distances \cite{walker_exciton_2013,rosenberg_electrically_2016}, and recently WG-polaritons operating up to room temperature have been reported\cite{ciers_propagating_2017}. Significant nonlinearity of waveguide polaritons has also been reported, leading to the formation of dark solitons in a WG-polariton fluid\cite{walker_dark_2017,walker_ultra-low-power_2015}. Since the WG-polaritons can be made in thin structures (since no thick cladding DBRs are required), it enables easy local access to both the optical modes, the photonic part (e.g. by patterning the WG surface), as well as to the matter, the excitonic part. In this regard, we showed that surface electrodes can be used to easily apply voltages across the QW planes, resulting in an electrical polarization of the excitons, that acquire an electrical dipole moment. This allows an electrical control of the polariton signal \cite{rosenberg_electrically_2016} and results in a significant electrically controlled enhancement of the mutual interactions between WG-polaritons\cite{rosenberg_strongly_2018,Togan2018StrongPolaritons}. Therefore, flying WG-polaritons can be excellent candidates for polariton-based circuitry, due to the simple access to local control of both the light and the matter parts of the polaritons, where a complex spatial control can be potentially designed and achieved by methods as simple as single-layer shallow etching or simple deposition processes.  
Until now, all experiments in WG-polaritons were done in a slab-WG geometry, with no lateral optical confinement, where polaritons are confined only in one dimension. This simple geometry limits the ability to construct complex lightwave structures, as well as the ability to maintain control of the direction and density of the WG-polaritons,  since polaritons in a slab geometry display a significant lateral spreading as they propagate.\\
In this paper we demonstrate two methods to confine polaritons laterally  in  channel-WG geometries. In the first part of this paper we show that polaritons can be confined in an etched rectangular WG (RWG) \cite{hunsperger_integrated_2009} where polaritons propagate to distances much larger than those possible in a slab WG, and where they can also propagate along a bent channel. We then show that the dielectric mismatch between a slab-WG and a thin dielectric strip is enough to create a low-loss strip-loaded polariton WG (SWG) leading to a very large increase of the propagation length. We demonstrate that a thin strip of ITO can be used to both optically guide polaritons and electrically control over the excitonic part of the optically confined polaritons by using it as an electrode which can electrically polarize the SWG-polaritons. 
\begin{figure}[tbp]
\includegraphics[width=0.5\textwidth]{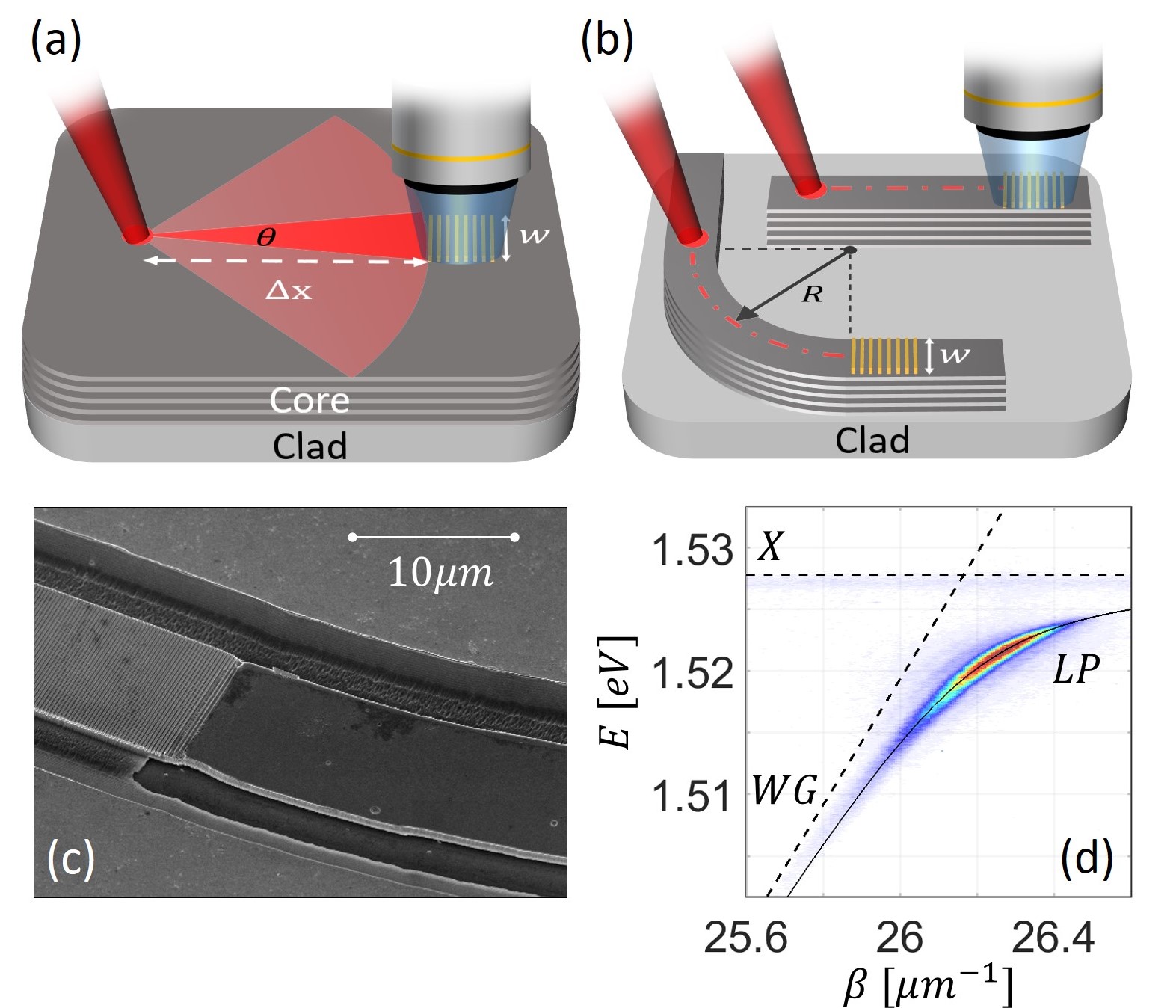}
\caption{\label{fig1}
(a) Slab-waveguide experiments: a non-resonant excitation beam is focused on a planar waveguide at a distance of $\Delta x$ away from the grating coupler to generate a cloud of polaritons that propagate isotropically. The fraction of these polaritons that propagate toward the grating coupler, illustrated by the $\theta$ red cone, are coupled out and collected by the collection objective. (b) Illustration of an RWG and a bent-RWG. The dashed dotted lines depict the line of the non-resonant excitation with varying  $\Delta x$. (c) An SEM image of a section of the bent-RWG, with the metallic grating at the left part of the channel. (d) PL measured dispersion of the planar WG polaritons emitted from the grating coupler. The bare exciton and photons modes are marked by dashed lines and the fitted LP branch is marked by a solid line.}
\end{figure}

\section{Results and discussion}

\subsection{A rectangular polariton waveguide}
\begin{figure*}[tb]
\includegraphics[width=0.95\textwidth]{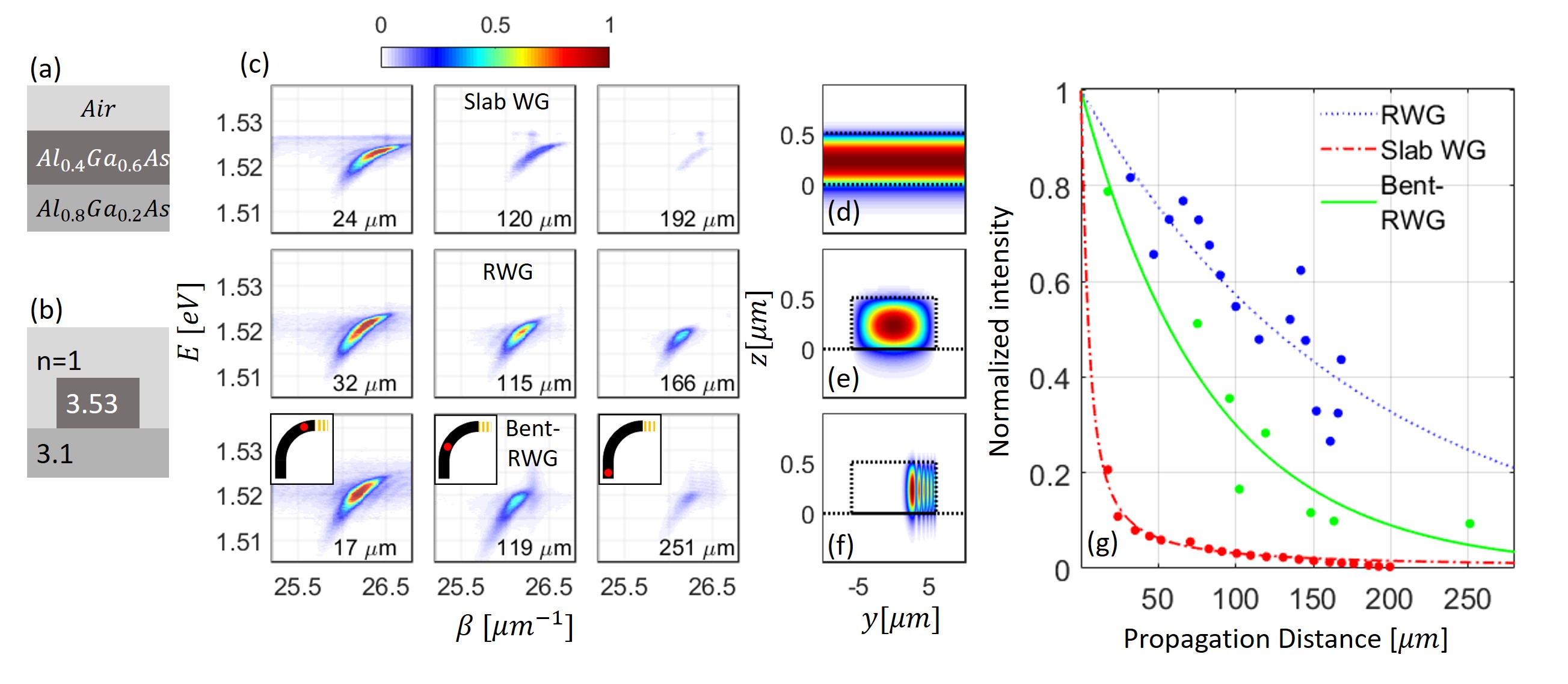}
\caption{\label{fig2} 
A comparison between the cross-sections of a slab-WG (a) and a RWG (b). (c) PL dispersion measurements: The different rows correspond to a slab-WG, a straight-RWG and a bent-RWG correspondingly. $\Delta x$ increases from left to right in each row. For the bent-RWG insets are added to illustrate the excitation position (red dot). (d-f) Calculated profiles of the fundamental WG modes in the slab-WG, the RWG and the bent-RWG respectively. The intensity in each panel is normalized to the intensity from the shortest $\Delta x$. (g) The variation in the emitted intensity as a function of $\Delta x$. The straight and bent RWG channels are fitted to exponential decay, a curve of $a\times \tan^{-1}(\frac{w}{2x})$ is plotted with $w=10\mu m$ as a guideline for the slab-WG measurements.} 
\end{figure*}

In this section we compare between the propagation lengths of polaritons in three different WG geometries, a slab WG, a 10 $\mu m$-wide straight RWG and a 10 $\mu m$-wide bent-RWG with a radius of curvature of 100 $\mu m$, all illustrated in Fig.\ref{fig1}(a,b). The sample, and the planar WG used in this work is identical to the one used by Rosenberg et. al.\cite{rosenberg_electrically_2016} and incorporates twelve 20nm wide GaAs/Al$_{0.4}$Ga$_{0.6}$As QWs in its core. Both RWGs were fabricated by etching two parallel $\sim 600  nm$-deep trenches resulting in a rectangular WG channel. A comparison between the cross-sections of the slab-WG and the RWG is shown in Fig. ~\ref{fig2}(a,b). On each RWG a 40 $\mu m$ long and 10 $\mu m$ wide metallic grating-coupler were deposited (period=240 nm, duty cycle=0.5) to allow out-coupling of the propagating polaritonic signal\cite{rosenberg_electrically_2016}. The method for measuring the propagation lengths of the polaritons is the following: a non-resonant excitation beam is focused at a specific distance $\Delta x$ from the grating-coupler, generating a reservoir of uncoupled excitons at the excitation spot. A large fraction of these excitons relaxes and accumulates at the bottleneck of the lower polariton (LP) dispersion and propagates inside the WG with a wave vector $\beta$. The fraction of these polaritons which propagates toward the grating-coupler is coupled out to be collected and angularly resolved by the acquisition setup. In Fig .\ref{fig1}(d) We show the spectrally and angularly resolved photoluminescence (PL) emitted from a slab WG through the grating coupler. The intensity of the PL is determined by the polariton density reaching the grating from excitation spot after traveling a distance of $\Delta x$. In Fig.\ref{fig2}(c)  we show examples of the PL dispersions measured when exciting at three locations along each of the WGs.
Here the different rows correspond to a slab-WG, a straight-RWG and a bent-RWG respectively, and the excitation distance $\Delta x$ increases from left to right. Several things can be learned from Fig.\ref{fig2}(c). In all these measurements, the dispersion of the LP can be clearly seen and the total intensity decreases when increasing $\Delta x$. The variation in the emitted intensity as a function of $\Delta x$, fitted to exponential decay for each of the RWG geometries, $I(\Delta x)=I_0\exp(-\Delta x/L)$ is shown in Fig.\ref{fig2}(g). It is apparent that the propagation length in the straight-RWG is much larger than that in a slab-WG. This can be explained when considering geometrical losses due to spatial spreading in a slab-WG due to the lack of lateral confinement: non-resonantly excited polaritons will propagate in all directions away from their excitation spot. We only collect PL from polaritons that passes under the grating-coupler, which propagate within an angle of $\theta=2\tan^{-1}(\frac{w/2}{\Delta x})$, where $w$ is the width of the grating-coupler as is illustrated in Fig.\ref{fig1}(a). Obviously, as we increase $\Delta x$, $\theta$ reduces and the measured intensity decreases. Good agreement between the above geometrical dependence and the experiment is seen in red in Fig.\ref{fig2}(g). Due to the lateral confinement in the case of RWG, the polaritons are guided towards the grating-coupler and are subjected only to physical losses which affect the polariton population. This difference can be understood when comparing the calculated optical modes cross-sections in each geometry. The fundamental optical modes for each of the WG geometries, calculated using a commercial finite difference eigenmode solver (Lumerical), are presented In Fig.\ref{fig2}(d-f). It can be seen that while the mode of the slab-WG is confined only along the growth direction, the modes of the RWGs are confined in two dimensions. Particularly, the calculations show that a mode-confinement is expected also in a bent-RWG geometry which suggests that polaritons can be guided along curved trajectories, a feature which is demonstrated experimentally in the bottom row of Fig.\ref{fig2}(c). Due to the lateral confinement of the RWG and the Bent-RWG we see an increased propagation length $L$ as shown in Fig.\ref{fig2}(g). We measured $L=179\pm60\mu m$ in the straight-RWG and $L=83\pm30\mu m$ in the bent-RWG. Most of the loss of the RWG originates from scattering of polaritons from the rough surface of the side walls of the RWG. These scatterings are especially dominant in the bent-RWG where the mode is squeezed to the side and has a higher overlap with the rough wall of the WG (see Fig.\ref{fig2}(f)) which explains the relatively small value of $L$ of the bent-RWG. The side walls of both RWG channels were created using a wet-etch process.  We note that optimization of the process and using dry-etch techniques will significantly reduce the roughness and should result in significantly larger propagation lengths..

\subsection{An electrically-active striploaded polariton waveguide}
\begin{figure}[tbp]
\includegraphics[width=0.5\textwidth]{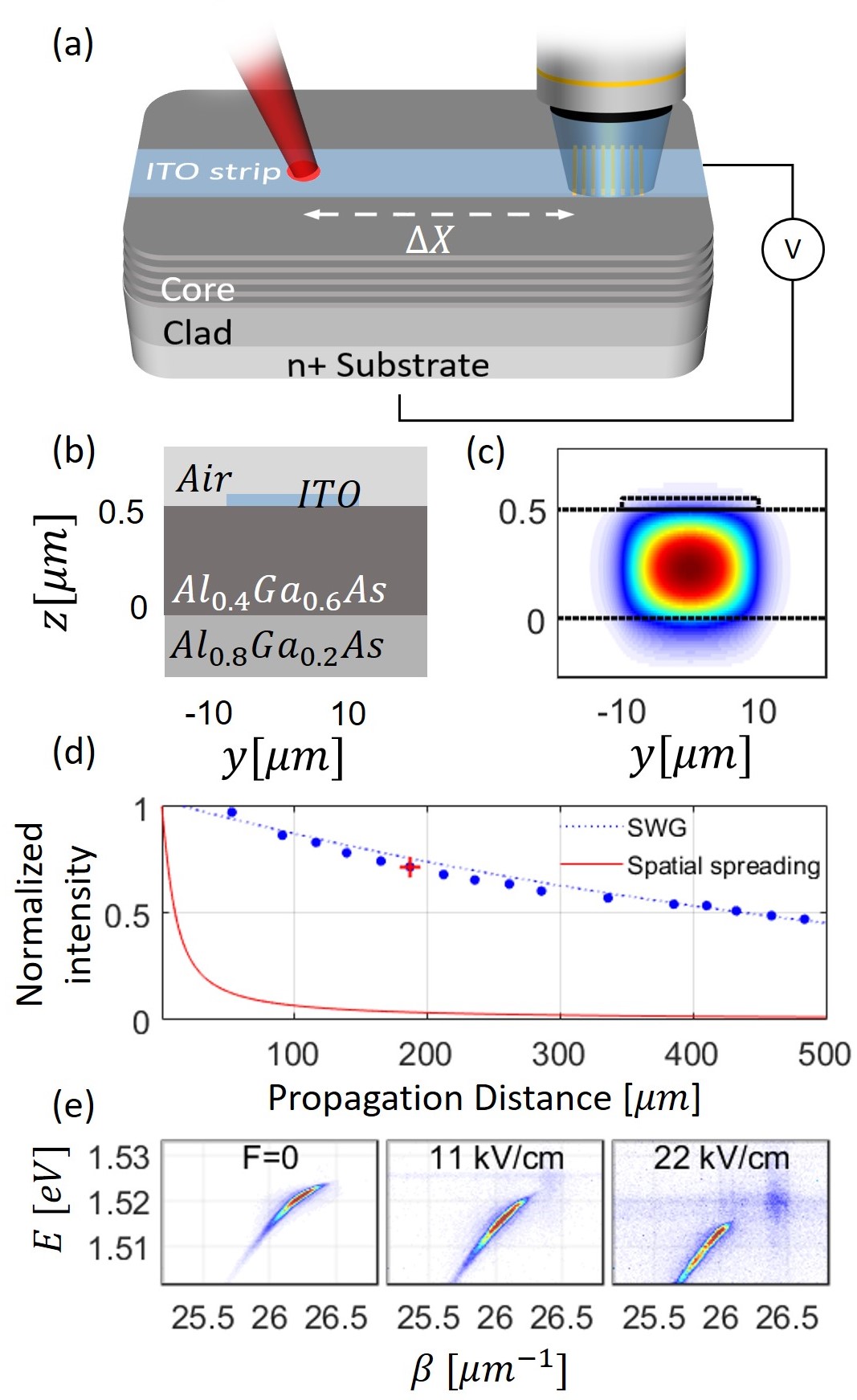}
\caption{(a) Illustration of an electrically biased SWG. (b,c) The cross section, and the calculated profile of the fundamental mode of the SWG. (d) The variation of the measured emitted intensity as a function of $\Delta x$ fitted to a decaying exponent (blue). The red line shows the calculated expected intensity in a slab-WG with only geometrical losses due to unconfined lateral spreading. (e) PL dispersions measured while inducing different values of applied electric field $F$. The Stark red-shift of the polariton signal is a signature of the formation of dipolaritons \cite{rosenberg_electrically_2016,rosenberg_strongly_2018}.\label{fig3}}
\end{figure}

Next we turn to present an electrically active striploaded-WG (SWG), where lateral optical confinement is achieved together with an electrical control over the QW-excitons composing the polaritons. The SWG was fabricated by laying a 50 nm thick, 20 $\mu m$ wide strip of ITO on top of the planar WG, as is shown in In Fig.\ref{fig3}(a,b). The  ITO strip introduces a higher effective refractive index than its surrounding, leading to a laterally confined WG mode, as is shown in the numerical calculation presented in Fig.\ref{fig3}(c).
The measured propagation decay of the polariton intensity along the SWG is presented in Fig.\ref{fig3}(d) where it is compared to a calculation of the expected geometrical losses due to spatial spreading in slab-WG with a 20 $\mu m $ wide grating-coupler. The fact that the propagation decay in the SWG is much slower than the calculated spatial spreading in slab-WG, confirms that the polaritons are indeed laterally confined by the SWG. Furthermore, we find a propagation length $L=610 \pm 50 \mu m$, longer than that of the RWG. This is attributed to the lack of etching, which significantly reduces roughness-induced optical losses. As far as we know, such long polariton propagation lengths, approaching a millimeter, have not been achieved in any other geometry.

Finally, we want to emphasize the added functionality that is easily introduced in the ITO-based SWG geometry. Here the conductive ITO strip also forms a transparent top electrode through which voltage can be applied across the sample. In Fig.\ref{fig3}(e) we plot three PL measurements measured when exciting at a distance of $187\mu m$ from the grating-coupler while applying different values of voltage across the sample with respect to the n$^+$ doped substrate. The effect of the Stark red-shift, induced by the electric field, on the dispersion can be clearly seen. In addition, the ability to measure the WG-polariton signal when exciting this far away from the grating-coupler, indicates that the effect of the electric field on the confinement is negligible. This dual use of the ITO strip allows therefore to form fully guided modes of electrically polarized WG-polaritons, with electrical control over the polariton energy, dispersion, and interactions as has been recently demonstrated \cite{rosenberg_electrically_2016,rosenberg_strongly_2018}.

\subsection{Summary and conclusions}
We demonstrated two types of WG geometries which can be used to laterally confine and guide polaritons. This lateral confinement maintains well-defined polariton modes over long propagation distances, prevents spreading and density reduction, and provides the ability to guide polaritons in curved trajectories.
These guiding capabilities are a necessary step towards the demonstration of large-scale on-chip polaritonic optical circuits. The demonstrated ability, to control simultaneously both the photonic part of the flying polaritons through optical confinement and the excitonic part using electrical fields, is a milestone towards a full control of polaritons.
Since electrically polarized polaritons exhibit stronger nonlinearities than unpolarized polaritons\cite{rosenberg_strongly_2018,Togan2018StrongPolaritons}, combined with the ability of strong optical confinement, the enhanced polariton interactions may allow observation of polariton-polariton interactions on the quantum level, which can open up routes towards polaritonic-based quantum gates.

\begin{acknowledgement}
The authors thank to financial support from the U.S. Department of Energy: Office of Basic Energy Sciences - Division of Materials Sciences and Engineering, from the United State - Israel Binational Science Foundation (BSF grant  2016112), and from the Israeli Science Foundation (grant No. 1319/12). The work at Princeton University was funded by the Gordon and Betty Moore Foundation through the EPiQS initiative Grant GBMF4420, and by the National Science Foundation MRSEC Grant DMR-1420541.
\end{acknowledgement}

\bibliography{bibi.bib}

\end{document}